\documentclass[5p,twocolumn,times,number]{elsarticle}
\usepackage[utf8]{inputenc}
\usepackage[english]{babel}
\usepackage{lineno,hyperref}
\usepackage{amsmath}
\usepackage{amsfonts}
\usepackage{amssymb}
\usepackage{upgreek}
\usepackage{subfigure}
\usepackage{ragged2e}
\usepackage{graphicx}
\graphicspath{ {Images/} }
\usepackage{txfonts}
\usepackage{multicol}
\usepackage{latexsym}

\def\pmbanner{{\hrule height 1 pt}\vskip35pt{NIMA POST-PROCESS BANNER TO BE REMOVED AFTER FINAL ACCEPTANCE}\vskip35pt{\hrule height 4pt}\vskip20pt} 

\begin{document}

\begin{frontmatter}

\title{\pmbanner Frequency domain multiplexing readout for large arrays of transition-edge sensors}
\author[a]{D.~Vaccaro}
\author[a]{H.~Akamatsu}
\author[a]{L.~Gottardi}
\author[b]{J.~van~der~Kuur}
\author[a]{K.~Nagayoshi}
\author[a]{E.~Taralli}
\author[a]{M.~de~Wit}
\author[a]{M.P.~Bruijn}
\author[a]{A.J.~van~der~Linden}
\author[a]{B.-J.~van~Leeuwen}
\author[a]{P.~van~der~Hulst}
\author[a]{K.~Ravensberg}
\author[a]{C.P.~de~Vries}
\author[c]{M.~Kiviranta}
\author[a,d]{J.-R.~Gao}
\author[a,e]{J.W.A.~den~Herder}

\address[a]{NWO-I/SRON Netherlands Institute for Space Research, 2333CA Leiden, The Netherlands}
\address[b]{NWO-I/SRON Netherlands Institute for Space Research, 9747AD Groningen, The Netherlands}
\address[c]{VTT, Tietotie 3, 02150 Espoo, Finland}
\address[d]{Optics Group, Delft University of Technology, 2628CJ Delft, The Netherlands}
\address[e]{Universiteit van Amsterdam, Science Park 904, 1090GE Amsterdam, The Netherlands}

\begin{abstract}

We report our most recent progress and demonstration of a frequency domain multiplexing (FDM) readout technology for transition-edge sensor (TES) arrays, both of which we have been developing, in the framework of the X-IFU instrument on board the future Athena X-ray telescope.  Using Ti/Au TES micro-calorimeters, high-Q LC filters and analog/digital electronics developed at SRON and low-noise two-stage SQUID amplifiers from VTT Finland, we demonstrated feasibility of our FDM readout technology, with the simultaneous readout of 37 pixels with an energy resolution at of 2.23 eV at an energy of 6 keV. We finally outline our plans for further scaling up and improving our technology.
\end{abstract}

\end{frontmatter}

\begin{quote}
\textbf{This paper has been submitted for publication in Nuclear Instruments and Methods A and is currently under review.}
\end{quote}

\section{Introduction}

Some of the future experiments pursuing scientific breakthroughs in the field of astronomy, cosmology or astroparticle physics will take advantage of the extreme sensitivities of cryogenic detectors such as transition-edge sensors (TES) \cite{tes}. A TES is a thin film of superconducting material weakly coupled to a thermal bath, typically at temperatures $T < 100$~mK, that can be used as a very sensitive thermometer by exploiting its sharp phase transition. TES-based microcalorimeters can be used as non-dispersive spectrometers, with desirable features such as high dynamic range, photon collecting area, quantum efficiency and energy resolution. We have been developing TES microcalorimeters for X-ray spectroscopy as backup option for the Athena X-ray Integral Field Unit (X-IFU) \cite{xifu,ken}, demonstrating under AC bias resolving power capabilities at a level of $E/\Delta E \simeq$~3000. TES are nowadays being fabricated in large arrays, at a level of hundreds or thousands of detectors in a single wafer batch. This makes necessary the developments of technologies allowing the readout of a large number of TES with a single readout channel, thus simplifying the harness scheme and thermal load at cryogenic stage. One of such readout scheme is the frequency domain multiplexing (FDM)~\cite{bbfb}, that we have developed in the framework of Athena X-IFU.

\section{FDM readout description and demonstration}

In the FDM scheme, the readout of a TES array is performed by placing a tuned high-$Q$ LC band-pass filter in series with each detector and providing an AC bias with an independent carrier in the MHz range. The signals of all the detectors in the readout chain are summed at the input coil of a Superconducting QUantum Interference Device (SQUID), which provides a first amplification at cryogenic temperature. Further amplification and conversion into digital is performed at room temperature by a control electronics board, which is also responsible for the bias carrier generation and demodulation of the output comb. We make use of a base-band feedback scheme, where the demodulated TES signals are again remodulated using the same carrier frequencies, compensated with a phase delay and fed back at the SQUID feedback coil, to null the current at the SQUID input: this allows a more efficient use of the SQUID dynamic range and effectively increases the number of pixels readable in a single readout chain.

In our FDM systems the bias frequencies range from 1~MHz to 5~MHz, with a channel separation of 100~kHz, allowing to read up to 40 detectors out of a single readout line. The inductance of the LC filters is fixed to $2~\upmu$H to keep the electrical bandwidth $~R/L$ constant, and the different frequencies are obtained varying the capacitance. Superconducting transformers connect the LC filters to the TES array to tune the current flowing through the detectors, which is summed at the input of a two-stage SQUID amplifier provided by VTT \cite{vtt}. Each TES is a Ti/Au bilayer with normal resistance $R_{\text{N}}$ at a level of 150~m$\Omega$ and critical temperature $T_{\text{C}} \simeq 90$~mK, coupled to an Au or Au/Bi absorber optimized for photon energies ranging from 1 keV to 10 keV. For our experiments we typically employ a $^{55}$Fe source generating 5.9~keV photons (Mn-K$\upalpha$ lines).

The setups are enclosed in a superconducting shield and hanged via Kevlar wires on the mixing chamber of a dilution refrigerator, operating at a bath temperature of 50~mK. The temperature is controlled via germanium thermistors and eventual remnant magnetic field is cancelled out via superconducting Helmoltz coils.

Using such setups we developed and demonstrated effective solutions to issues intrinsic to AC readout that at early stages used to spoil TES performance in multiplexing, $e.g.$ high-aspect ratio TES to mitigate non-linear Josephson effects \cite{lgjosephson} and a Frequency Shift Algorithm (FSA) to limit the impact of intermodulation distortions \cite{fsadavide}. We furthermore demonstrated excellent single-pixel spectral performance under AC readout \cite{martinhar}, thermal crosstalk \cite{xtalk} and environmental susceptibility levels, in particular to magnetic field \cite{suscep,martinB}.

\begin{figure}[!h]
\begin{center}
\includegraphics[width=0.48\textwidth]{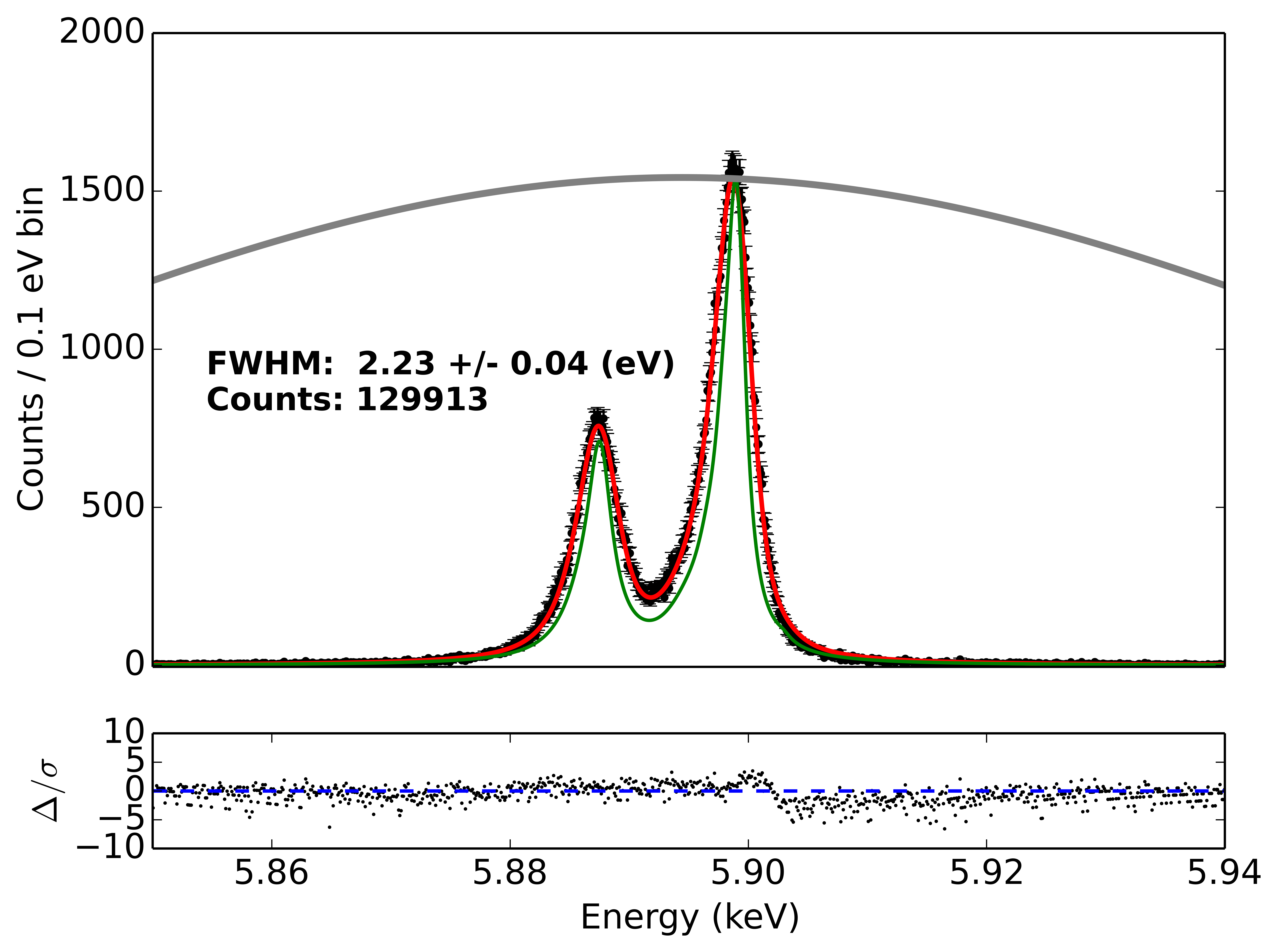}
\caption{Energy spectrum obtained under FDM readout of 37 TES, showing the summed energy resolution of the detectors for photons at Mn-K$\upalpha$ lines energies. Each photon energy is estimated using the optimal filter technique \cite{optfit}. Black dots are the binned data, red curve is the best fit from the Holzer model (green curve)~\cite{holzer} convolved with the energy resolution, obtained minimizing the Cash statistics \cite{cash} under the maximum likelihood method. They grey curve shows a comparison with the typical energy resolution ($\approx$~150~eV) of traditional solid-state detectors, such as CCD cameras.}\label{37mux}
\end{center}
\end{figure}

Taking on such optimizations allowed us to push the performance of our FDM readout, for which we recently demonstrated a spectral performance with 37 pixels of 2.2~eV at 5.9~keV, or equivalently a resolution of $\approx$~0.04\%, as shown in Figure~\ref{37mux}. More details on the FDM setups and demonstration can be found in \cite{hirokifdm}.

\section{Future outlook}

\begin{figure}[!h]
\begin{center}
\includegraphics[width=0.48\textwidth]{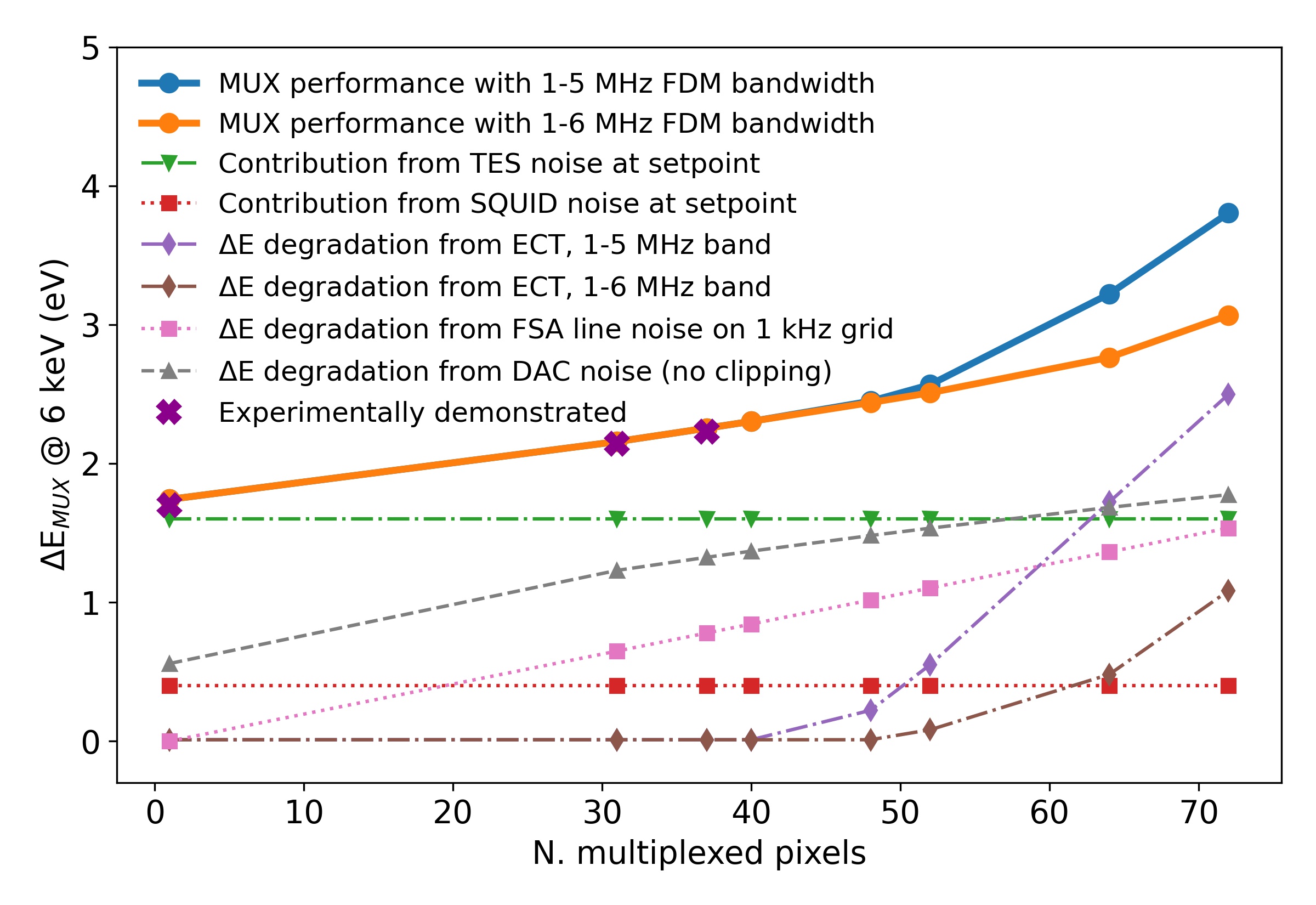}
\caption{Expected multiplexing performance as a function of multiplexing factor. The individual contributions (in quadrature) are considered either from an energy resolution budget compiled in the framework of X-IFU and from experimental data.}\label{nMUX}
\end{center}
\end{figure}

Our TES-based FDM readout technology does already satisfy the requirements for space-borne applications such as Athena X-IFU, nevertheless we envisage the possibility for further scaling up, as shown in Figure~\ref{nMUX}. For example, it appears to be feasible to expand the readout bandwidth up to 6~MHz, which would naturally allow to accommodate more pixels without a larger impact of electrical crosstalk (ECT). Multiplexing a larger number of pixels represents a challenge also in terms of DAC noise and intermodulation distortions contribution, which we plan to assess once new lithographic components will be fabricated in the near future.

Further activities currently in progress are tests with multi-channel FDM readout and measurements with a larger number of X-ray lines to verify with precision the calibrability of the TES energy scale. This activities will further validate the viability of our TES-FDM technology not only for astronomy but also for other possible applications, $e.g.$ plasma spectroscopy at fusion reactors \cite{iter} and solar axion research \cite{bgelba}.

\section*{Acknowledgements}

SRON is financially supported by the Nederlandse Organisatie voor Wetenschappelijk Onderzoek.

This work is part of the research programme Athena with project number 184.034.002, which is (partially) financed by the Dutch Research Council (NWO).

The SRON TES arrays used for the measurements reported in this paper is developed in the framework of the ESA/CTP grant ITT AO/1-7947/14/NL/BW.

\end{document}